\documentstyle[prl,aps,multicol,epsf]{revtex}
\begin{document}
\draft
\title{Abrupt Transition between 
Thermally-Activated Relaxation and 
Quantum Tunneling in a Molecular Magnet.}
\author{K. M. Mertes, Y. Zhong, and M. P. Sarachik}
\address{Physics Department, City College of the City University of New York,
New York, NY 10031}
\author{Y. Paltiel, H. Shtrikman, and E. Zeldov}
\address{Department of Condensed Mater Physics, 
The Weizmann Institute of Science, Rehovot 76100, Israel}
\author{E. Rumberger and D. N. Hendrickson}
\address{Department of Chemistry and
Biochemistry, University of California at San Diego, La Jolla, CA 92093}
\date{\today}
\maketitle \begin{abstract}
We report Hall sensor measurements of the magnetic relaxation of Mn$_{12}$-acetate 
as a function of magnetic field applied along the easy axis of magnetization.  
Data taken at a series of closely-spaced temperatures between $0.24$ K and $1.4$ K 
provide strong new evidence for an abrupt 
``first-order'' transition between 
thermally-assisted relaxation and magnetic decay via quantum tunneling.  

\end{abstract}
\vspace {6mm}
\begin{multicols}{2}

Single-molecule magnets are organic materials which contain a large (Avogadro's) 
number of identical magnetic molecules; 
([Mn$_{12}$O$_{12}$(CH$_3$COO)$_{16}$(H$_2$O)$_4$]$\cdot$ 2CH$_3$COOH$\cdot4$H$_2$O), 
generally referred to as Mn$_{12}$ acetate, is a particularly simple 
and much-studied example of this class.  The Mn$_{12}$ clusters are each composed 
of twelve Mn atoms (see Fig. 1) coupled by superexchange through oxygen 
bridges to give a sizable $S=10$ spin magnetic moment that is stable at temperatures 
of the order of $10$ K and below\cite{sessoli}.  These identical weakly-interacting 
magnetic clusters are regularly arranged on a tetragonal crystal.  As illustrated by 
the double well potential of Fig. 1, shown there in the presence of a longitudinal 
field, strong uniaxial anisotropy (of the order of 
$65$ K) yields doubly degenerate ground states in zero field and a set of excited 
levels corresponding to different projections $m_s = \pm 10, \pm 9,.....,0$ of the 
total spin along the easy c-axis of the crystal.  Measurements\cite{friedman,thomas} 
below the blocking temperature of 
$3$ K have revealed a series of steep steps in the curves of $M$ versus $H$ at 
roughly equal 
intervals of magnetic field, as shown in Fig. 2, due to enhanced relaxation 
of the magnetization whenever levels on opposite sides of the anisotropy barrier 
coincide in energy.  As demonstrated by the data of Fig. 2, different ``steps'' 
dominate at different temperatures, indicating 
that thermal processes play a central role.  The steps in the magnetization curves 
have been attributed\cite{novak} to thermally-assisted quantum tunneling of the 
spin magnetization.

Thermally-assisted tunneling is shown schematically for the third ``step'' or 
field-resonance by the sequence of straight-line 
arrows in Fig. 1: the magnetization is 
thermally activated to a level near the top of the 
metastable well ({\it e. g.} $m'=-5$), tunnels across the barrier (to 
$m=2$), and decays to the ground state ($m=10$) of the stable well.  Thermal 
activation becomes exponentially more difficult as one 
proceeds up the ladder to higher energy levels; on the other hand, the barrier 
is lower and more penetrable, so that the tunneling process becomes 
exponentially easier.  Which level (or group of adjacent levels) dominates the 
tunneling is determined by competition between the two effects.  
As the temperature is reduced and thermal activation becomes more difficult, the 
states that are active in the tunneling move gradually to lower energies 
deeper in the potential well.

\vbox{
\vspace{0.0in}
\hbox{
\hspace{-0.2in} 
\epsfxsize 3.4in \epsfbox{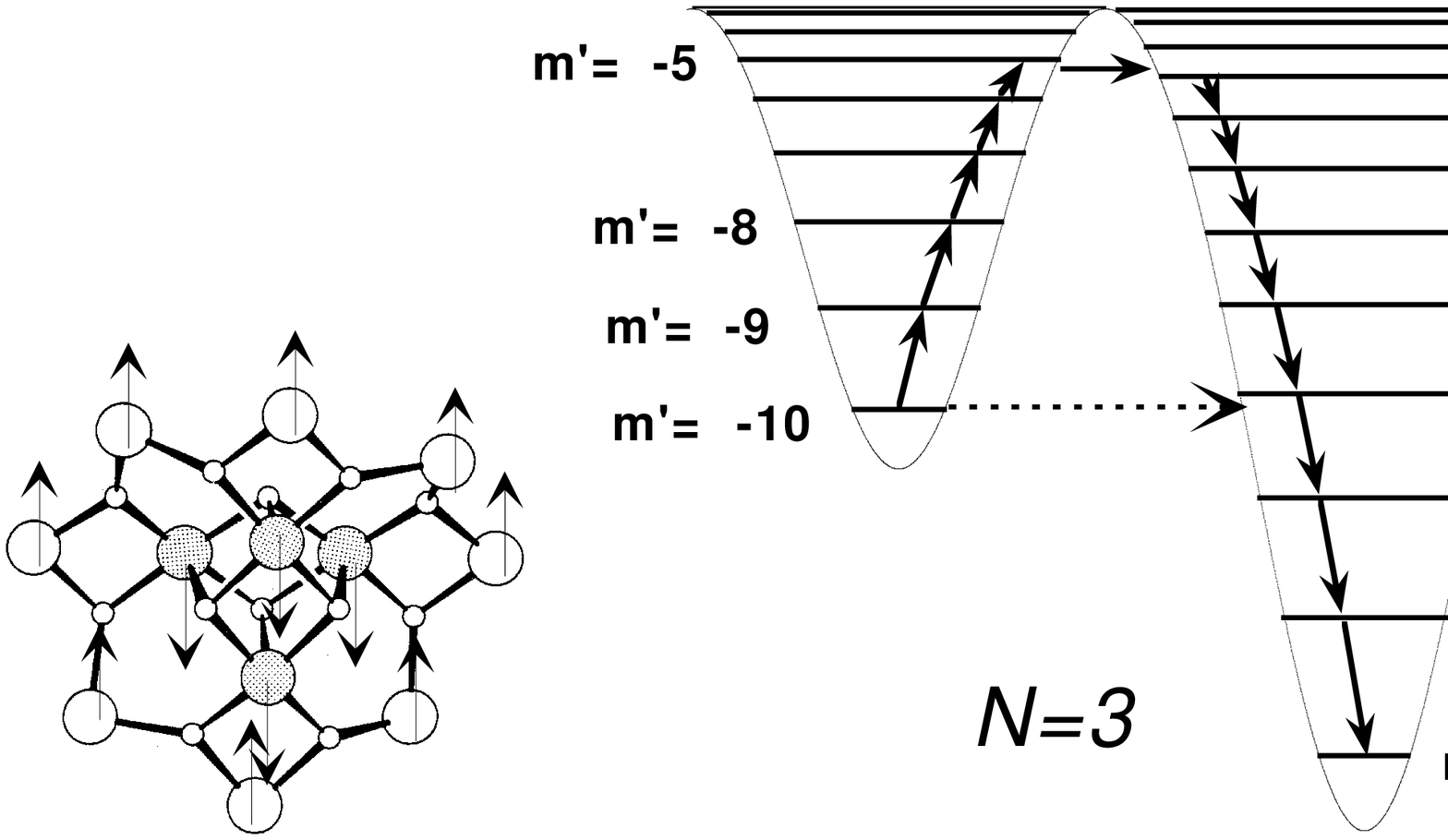} 
}
}
\refstepcounter{figure}
\parbox[b]{3.3in}{\baselineskip=12pt FIG.~\thefigure.
Double-well potential in the presence of a longitudinal magnetic field applied 
along the easy c-axis.  Thermally-assisted tunneling is indicated by straight-line 
arrows.  The left side of the figure shows a schematic of the 
Mn$_{12}$ molecule composed of four inner spin 
($S=-3/2$) Mn$^{4+}$ ions and eight outer spin ($S=+2$) Mn$^{3+}$ ions with 
oxygen bridges, 
yielding a total spin $S=10$ ground state at low temperatures.
\vspace{0.13in}
}
\label{1}

\vbox{
\vspace{0.2in}
\hbox{
\hspace{0.15in} 
\epsfxsize 2.5in \epsfbox{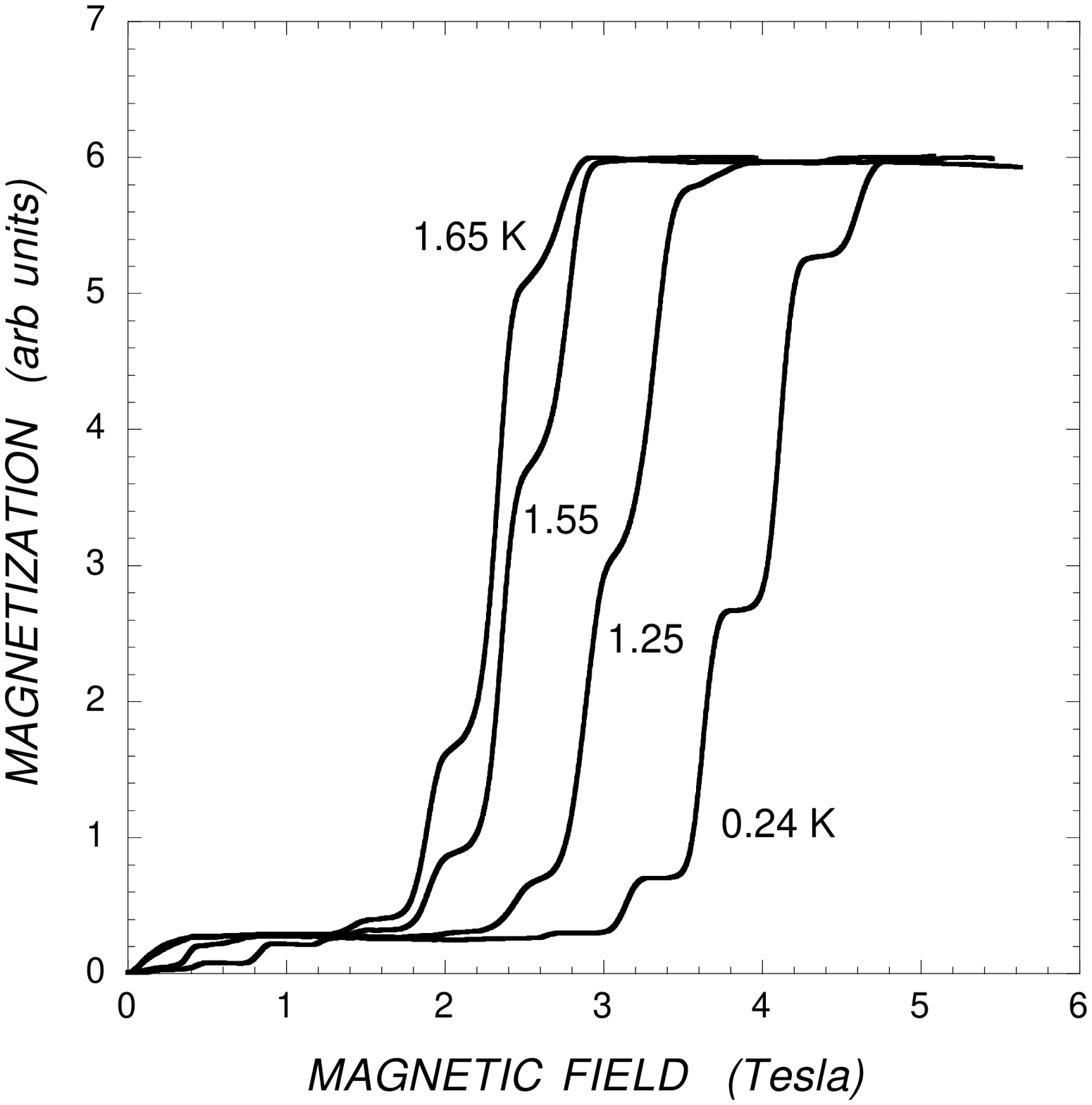} 
}
}
\refstepcounter{figure}
\parbox[b]{3.3in}{\baselineskip=12pt FIG.~\thefigure.
Magnetization versus longitudinal magnetic field for a Mn$_{12}$ sample 
starting from a demagnetized state, $M=0$; data are shown at four different 
temperatures, as labeled.  
Note the steep segments, or steps, corresponding to faster magnetic relaxation 
at specific values of magnetic field.
\vspace{0.10in}
}
\label{2}

Chudnovsky and Garanin\cite{firstorder} have recently proposed that as the 
temperature is reduced, the levels that dominate the tunneling can shift to 
lower energies either continuously (``second order'' transition) or abruptly (``first 
order'' transition), depending on the form of the potential.  It may thus 
be possible to observe an abrupt transition from thermally assisted tunneling 
(straight-line arrows in Fig. 1) to pure quantum mechanical tunneling 
from the lowest state of the metastable well (denoted by the dotted line) as the 
temperature is reduced.  Indeed, earlier magnetization experiments\cite{kent} have 
indicated that a rapid shift occurs to tunneling from the lowest state as the 
temperature is reduced.

In the present paper we report detailed measurements at a series of very closely 
spaced temperatures of the magnetization as a function of magnetic field applied 
along the easy axis of magnetization.  We show that there is an abrupt transfer 
over a narrow range of temperature to enhanced magnetic relaxation at 
the magnetic field corresponding to tunneling from the lowest state of the 
metastable well.  These results provide solid new evidence that there is an abrupt 
or ``first order'' transition between thermally-assisted and ground state tunneling 
in Mn$_{12}$.

Identification of the levels that participate in tunneling is based on the 
following considerations.  The spin Hamiltonian for Mn$_{12}$ is given by:
\begin{equation}
{\cal H} = -D S_z^2 -g_z \mu_B H_z S_z - A S_z^4 + .......
\end{equation}
where $D$ is the anisotropy, the second term is the Zeeman energy, 
and the third on the right-hand side represents the next higher-order term in 
longitudinal anisotropy; additional contributions (transverse internal magnetic 
fields, transverse anisotropy,...) are not explicitly shown.  Tunneling occurs 
from level $m'$ in the metastable well to level $m$ in the stable potential 
well for magnetic fields:
\begin{equation}
H_z=N\frac{D}{g_z  \mu_B}\left[1 + \frac{A}{D}\left(m^2 +
m'^2\right)\right],
\end{equation}
where $N = |m + m'|$ is the step number.  The second term in brackets is small 
compared to $1$.  Thus, a series of steps $N_i$ occurs at approximately equally 
spaced intervals of magnetic field, $D/(g_z\mu_B) \approx 0.42$ Tesla; for a given 
step all pairs of levels cross at roughly the same magnetic field.  
However, careful measurements show that there is structure within each step due 
to the presence of the term $AS_z^4$; as shown diagrammatically in Fig. 3, 
the levels do not cross simultaneously, an effect that is more pronounced for 
levels that are deeper in the well.  EPR\cite{barra} and neutron 
scattering\cite{zhong,hennion,mirebeau} experiments have yielded precise values of 
$A=1.173(4)\times 10^{-3}$ K/mol and $D= 0.548(3) K$, and an estimate for 
$g_z$  of $1.94(1)$.  Comparison of the measured magnetic fields with those 
calculated from Eq. (2) therefore provides an experimental tool that allows 
identification of the states that are predominantly responsible for the tunneling.

\vbox{
\vspace{0.2in}
\hbox{
\hspace{0.0in} 
\epsfxsize 3.0in \epsfbox{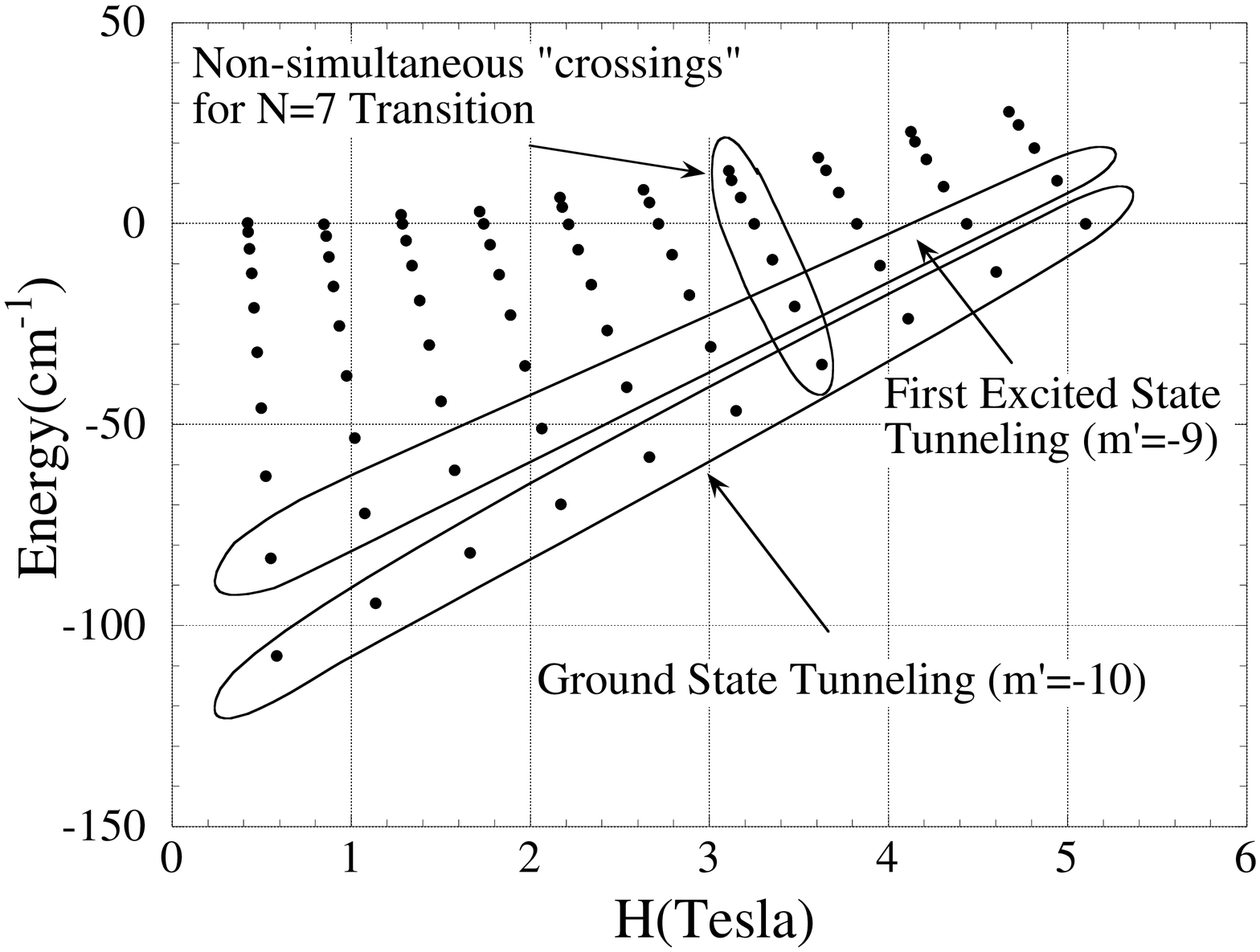} 
}
}
\refstepcounter{figure}
\parbox[b]{3.3in}{\baselineskip=12pt FIG.~\thefigure.
Energy level diagram obtained for the spin Hamiltonian for Mn$_{12}$.  
The dots indicate magnetic fields where pairs of energy levels on opposite sides of
the barrier cross, giving rise to enhanced magnetic relaxation via tunneling.}
\vspace{0.10in}

\label{3}

The magnetization of small single crystals of Mn$_{12}$-acetate was determined from 
measurements of the local magnetic induction at the sample surface using 
$10 \times 10$ $\mu$m$^2$ 
Hall sensors composed of a two-dimensional electron gas (2DEG) in a GaAs/AlGaAs 
heterostructure.  The 2DEG was aligned parallel to the external magnetic field, and 
the Hall sensor was used to detect the perpendicular component (only) of the magnetic 
field arising from the sample magnetization\cite{zeldov}.  

Our results are shown in the next few figures.  For different temperatures 
between $0.24$ K and $0.88$ K, Fig. 4 shows the first derivative, 
$\partial M/\partial H$, of the magnetization $M$ with respect to the 
externally applied magnetic field $H$\cite{corrections}.  The maxima occur at 
magnetic fields corresponding to faster magnetic relaxation due to level crossings 
on opposite sides of the anisotropy barrier.  In the temperature range of these 
measurements, maxima are observed for $N=|m+m'| =5$ through $9$.  Considerable 
structure associated with different pairs $m, m'$ is clearly seen within each 
step $N$, with a transfer of ``spectral weight'' to higher values of $m'$ deeper in 
the well as the temperature is reduced.  The 
issue is whether this transfer occurs gradually or abruptly.

\vbox{
\vspace{0.2in}
\hbox{
\hspace{-0.2in} 
\epsfxsize 3.2in \epsfbox{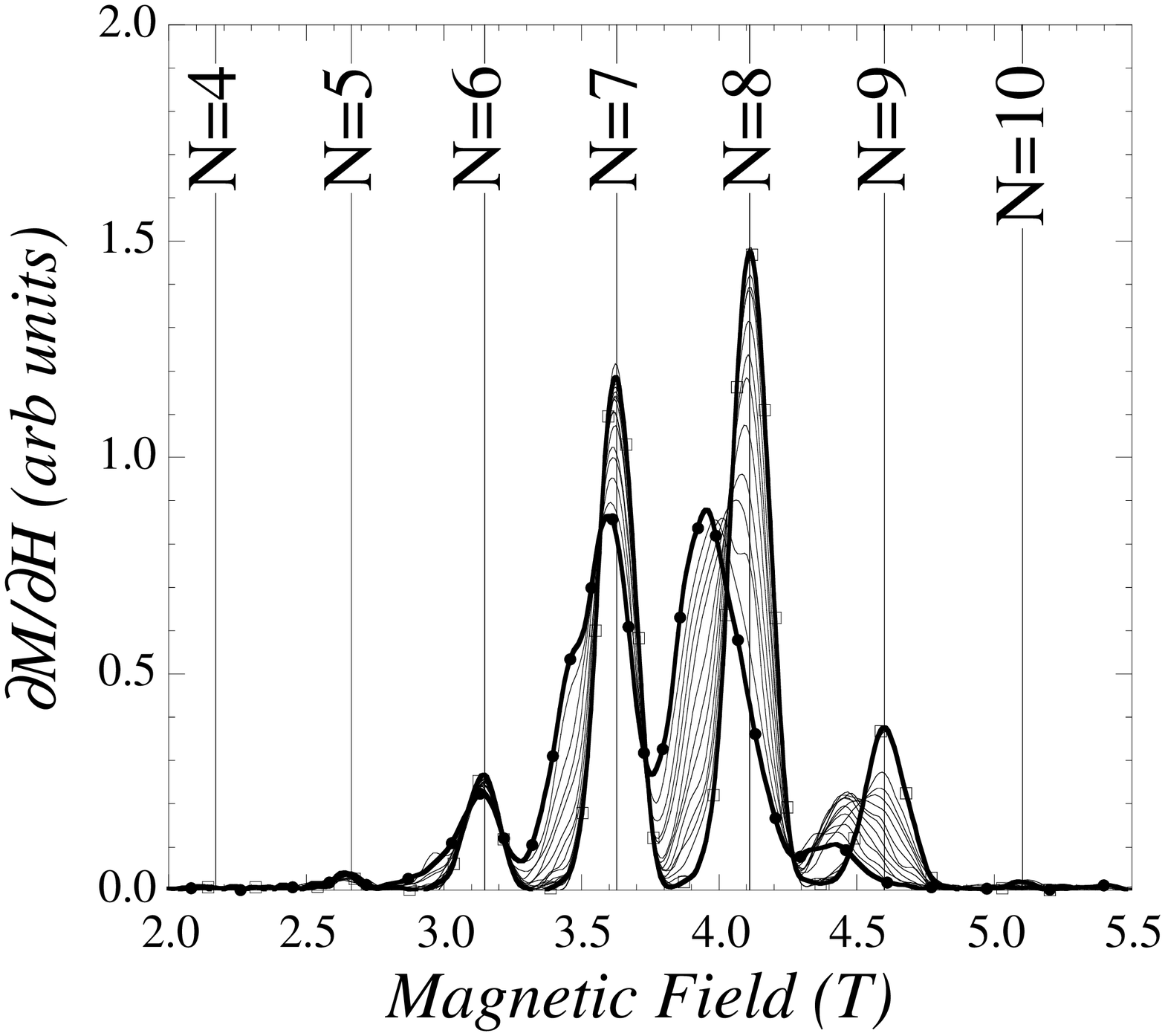} 
}
}
\refstepcounter{figure}
\parbox[b]{3.3in}{\baselineskip=12pt FIG.~\thefigure.
For a set of closely spaced temperatures, $\partial M/\partial H$ is shown 
as a function of 
magnetic field.  The amplitude is a measure of the rate of magnetic relaxation.  
Note the substructure within each of the four maxima corresponding to steps 
$N= |m'+m|=5, 6, 7, 8,$ and $9$. 
\vspace{0.10in}
}
\label{4}

In order to address this question, we examine some of the data in greater 
detail.  The derivative of the magnetization is shown on an expanded scale for step 
$N=7$ in Figs. 5 and 6.  Figure 5 shows $\partial M/\partial H$ as a function of 
magnetic field for different temperatures between $0.24$~K and $1.32$~K; 
the vertical lines indicate tunneling from levels corresponding to the different 
spin projections $m'$.  Figure 6 shows the same data in the H-T plane, with  
$\partial M/\partial H$ shown in the third dimension by different shading, with 
lighter shade corresponding to larger amplitude.  As the temperature is reduced, the 
maximum gradually moves to higher field and its amplitude changes.  Figure 5 shows 
that there is structure at some 
temperatures that indicates the presence of more than one maximum, implying 
that more than one pair of levels is active; where a single maximum appears, it is 
probably the convolution of two or three maxima.  It is noteworthy that the 
contribution from $m'=-9$ is minimal, or quite small, compared with other levels.  
In contrast, the contribution from $m'=-10$ becomes increasingly dominant as the 
temperature is lowered.  There is an 
abrupt transfer of weight to tunneling from the lowest (ground) state of the 
metastable well.

\vbox{
\vspace{0.2in}
\hbox{
\hspace{-0.2in} 
\epsfxsize 3.4in \epsfbox{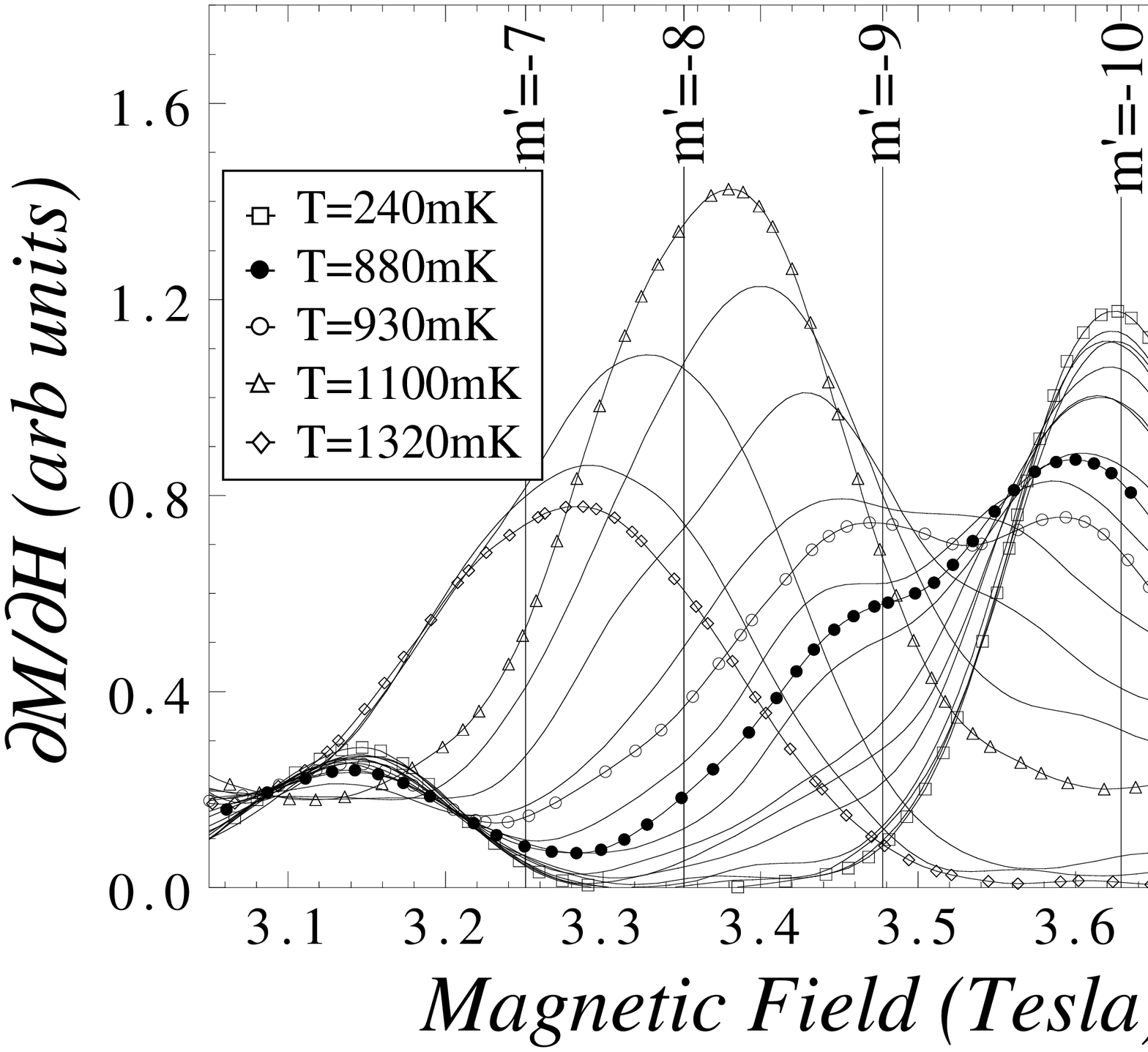} 
}
}
\refstepcounter{figure}
\parbox[b]{3.3in}{\baselineskip=12pt FIG.~\thefigure.
The first derivative of the magnetization with respect to magnetic
field versus magnetic field shown on an expanded scale for $N=|m'+m|=7$.  
The vertical lines denote the magnetic fields corresponding to tunneling between 
different pairs of levels (m',m) on opposite sides of the potential barrier: 
(-7,0), (-8,1), (-9,2), and (-10,3).  Several intermediate temperatures 
were omitted from the legend for clarity.}
\vspace{0.1in}

\label{5}

\vbox{
\vspace{0.3in}
\hbox{
\hspace{0.1in} 
\epsfxsize 3in \epsfbox{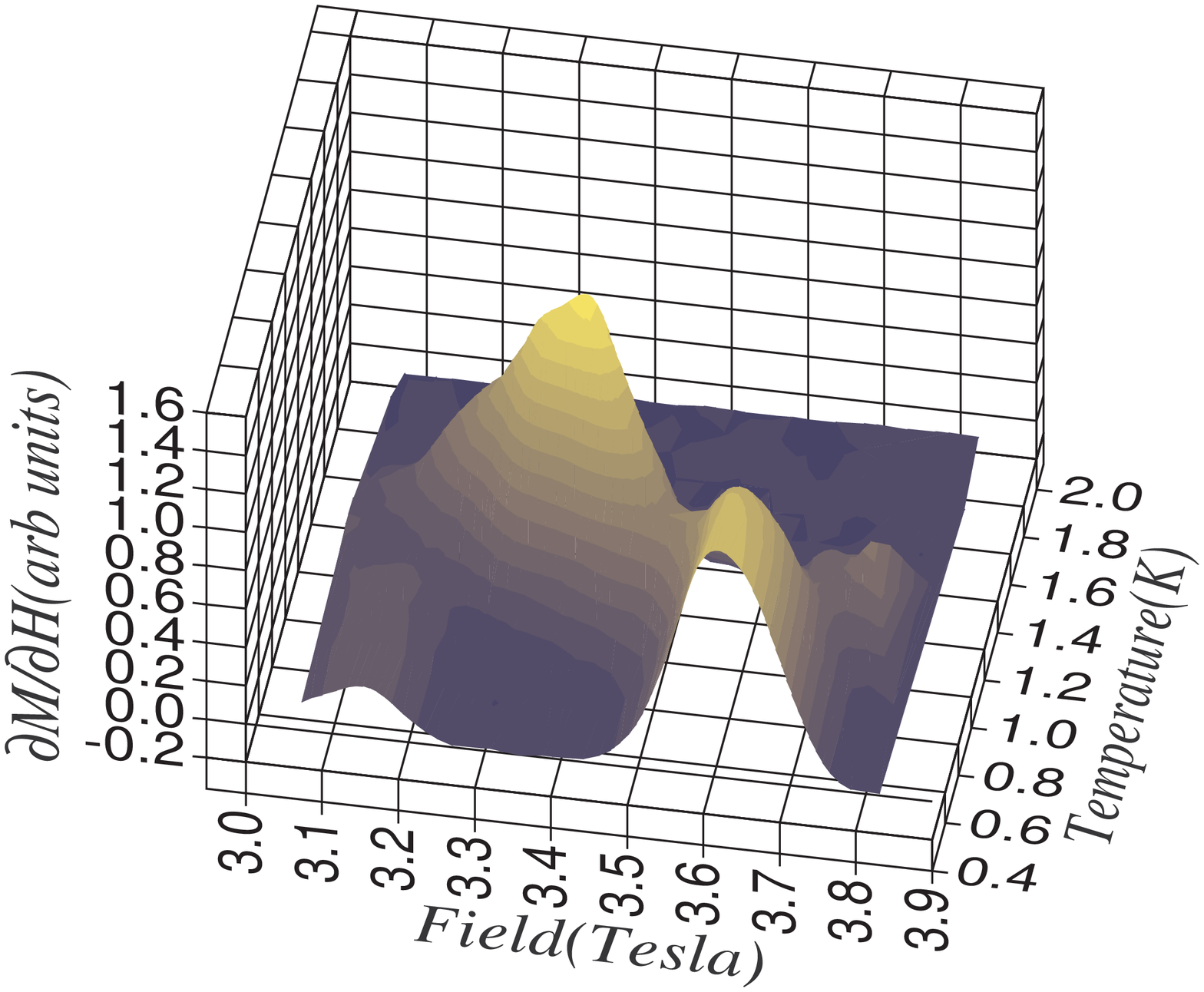} 
}
}
\vspace{0.2in}

\refstepcounter{figure}
\parbox[b]{3.3in}{\baselineskip=12pt FIG.~\thefigure.
For step $N=|m'+m|=7$, $\partial M/\partial H$ is shown as a function of 
magnetic field H and temperature T.  Yellow shading denotes large 
$\partial M/\partial H$ while blue corresponds to smaller amplitudes.}
\vspace{0.1in}

\label{7}

\vbox{
\vspace{0.in}
\hbox{
\hspace{-0.2in} 
\epsfxsize 3.3in \epsfbox{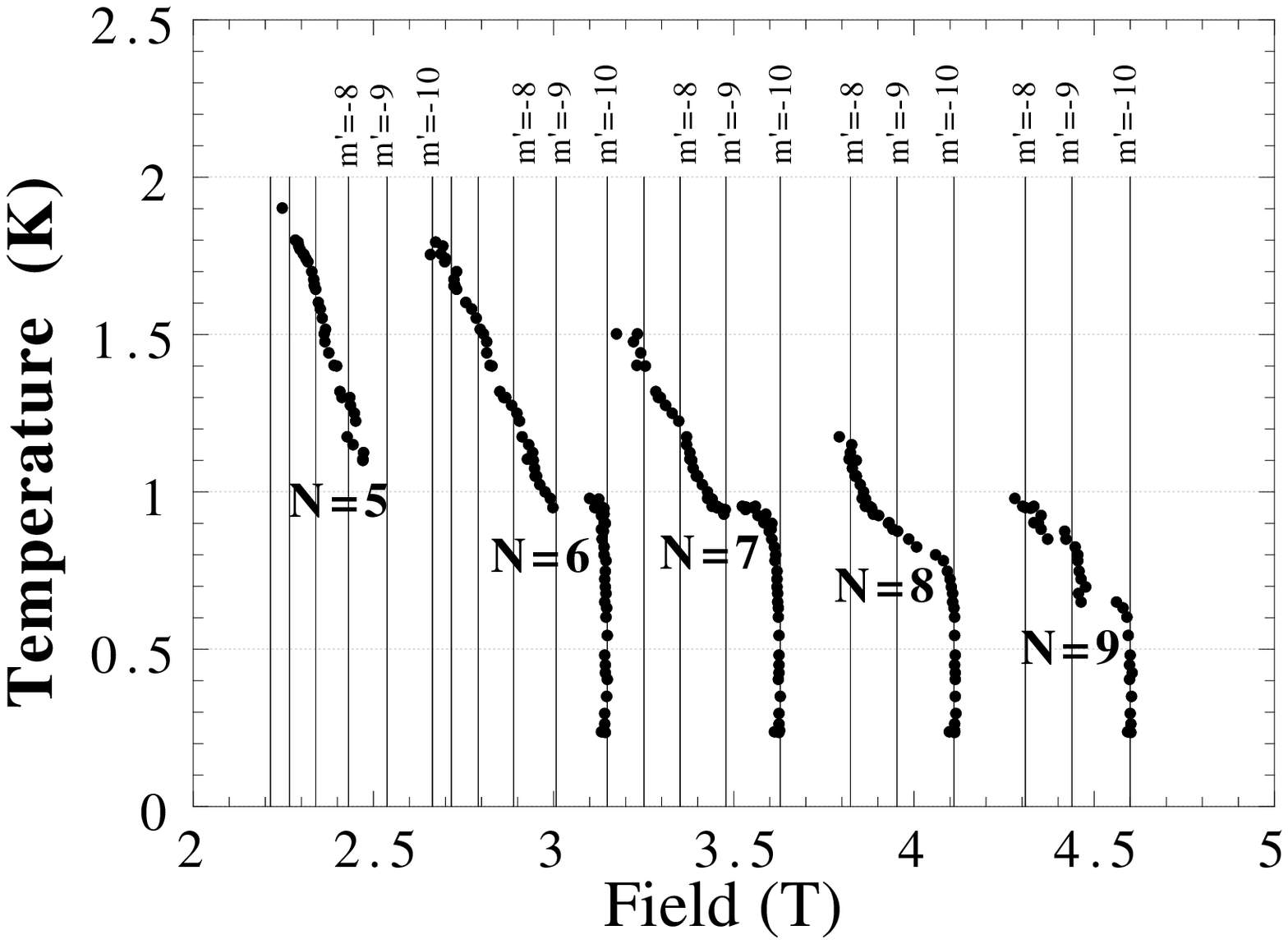} 
}
}
\refstepcounter{figure}
\parbox[b]{3.3in}{\baselineskip=12pt FIG.~\thefigure.
The magnetic field of the maxima in $\partial M/\partial H$ corresponding to
enhanced magnetic relaxation plotted as a function of temperature.  
The fields corresponding to tunneling from different levels $m'$ within 
each step $N$ are indicated by horizontal lines.  With decreasing temperature, 
the maxima initially shift gradually upward in field, then exhibit
an abrupt shift to tunneling from $m'=-10$ within a narrow temperature range,
below which the field of the maximum remains constant.}
\vspace{0.20in}

\label{7}

This is shown more explicity in Fig. 7.  Here, the positions of the maxima 
are plotted as a function of temperature for all measured $N$.  Within each 
step, the magnetic fields corresponding to tunneling from levels $m'$ in the 
metastable well are indicated by horizontal lines, as labeled.  
For each step, the position of the maximum shifts gradually and continuously 
to higher magnetic field as the temperature is decreased, and then moves 
abruptly to $m'=-10$ at some temperature below which the field of the maximum 
remains constant.  Although no levels are skipped entirely for the conditions 
of our experiments, there is a sudden shift to tunneling at a magnetic field that is 
independent of the temperature, and the value of this magnetic field agrees 
quantitatively with the calculated position for tunneling 
from $m'=-10$.

To summarize, magnetization measurements in Mn$_{12}$ acetate taken at closely 
spaced intervals of temperature exhibit an abrupt shift over a narrow 
temperature range to rapid magnetic relaxation at a ``resonant'' magnetic 
field corresponding to tunneling from the lowest state of the metastable potential 
well; this resonant field  is then independent of temperature as the temperature 
is reduced further.  Our data provide solid evidence for an abrupt transition 
between thermally-assisted tunneling and pure quantum tunneling 
from the ground state.

Work at City College was supported by NSF grant DMR-9704309 and at the 
University of California, San Diego by NSF grant DMR-9729339.  EZ acknowledges 
the support of the German-Israeli Foundation for Scientific Research 
and Development.

\end{multicols}

\begin{references}
\bibitem{corrections} Throughout this paper, we have used $H$ instead of 
the effective field $H_{eff} = H + \alpha (4\pi M)$; the field due to the sample 
magnetization is on the order of $300$ Oe.
\end{references}
\end{document}